\documentclass[10pt, twocolumn, twoside, letterpaper, journal, final]{IEEEtran}
\usepackage[noadjust]{cite}
\usepackage[utf8]{inputenc}
\usepackage{url}
\usepackage{footnote}
\usepackage{xcolor}

\usepackage[super]{nth}
\usepackage{longtable}
\usepackage{hhline}

\ifCLASSINFOpdf
  \else
    \usepackage[dvips]{graphicx}
     \DeclareGraphicsExtensions{.eps}
\fi
\usepackage{graphicx}
\usepackage[english]{babel}

\usepackage[framemethod=TikZ]{mdframed}

\mdfdefinestyle{MyFrame}{
    linecolor=black,
    outerlinewidth=1pt,
    roundcorner=6pt,
    innertopmargin=6pt,
    innerbottommargin=6pt,
    innerrightmargin=6pt,
    innerleftmargin=6pt,
    backgroundcolor=white}
\usepackage[framemethod=TikZ]{mdframed}

\begin{document}

\title{{\textbf{Context-Aware Security for 6G Wireless \\
\textit{The Role of Physical Layer Security}}}}
\vspace{3 cm}
\author{\textbf{\IEEEauthorblockN{ Arsenia Chorti$^1$, André Noll Barreto$^2$, Stefan Köpsell$^2$, Marco Zoli$^3$, Marwa Chafii$^{4}$, Philippe Sehier$^5$,   Gerhard Fettweis$^2$, H. Vincent Poor$^6$}}\\
$^1$ETIS UMR8051, CY Université, ENSEA, CNRS, F-95000, France\\
$^2$Barkhausen Institut gGmbH, Dresden 01187, Germany\\
$^3$Xilinx Dresden GmbH, Dresden, Saxony, DE\\
$^4$ Engineering Division, New York University (NYU) Abu Dhabi, 129188, UAE and NYU WIRELESS, NYU Tandon School of Engineering, Brooklyn, 11201, NY\\
$^5$Standardization Department, Nokia Bell Labs, Saclay, France\\
$^6$School of Engineering and Applied Science, Princeton University, Princeton, NJ, 08544}
\maketitle

\begin{abstract}
Sixth generation systems are expected to face new security challenges, while opening up new frontiers towards context awareness in the wireless edge. The workhorse behind  this projected technological leap will be a whole new set of sensing capabilities predicted for 6G devices, in addition to edge and device embedded intelligence. The combination of these enhanced traits can give rise to a new breed of adaptive and context-aware security protocols, following the quality of security (QoSec) paradigm. In this framework, physical layer security solutions emerge as competitive candidates for low complexity, low-delay and low-footprint, adaptive, flexible and context aware security schemes, leveraging the physical layer and introducing security controls across all layers, for the first time. 
\end{abstract}

\section{Introduction}
An intense discussion is currently underway with respect to the resilience and trustworthiness of the 6G radio, pivoting the enhancement of the security of the envisioned sixth generation (6G) wireless access. 
Notably, some of the recent, increasingly sophisticated attacks on the wireless edge, e.g., jamming or false base stations, can be implemented with a price tag as low as 1k$\$$ using low-cost software defined radios. In addition, we experience an expansion of the attack surface with artificial intelligence (AI) and machine learning (ML) tools. In parallel, 
as we move gradually away from the standard client-server networking paradigm and enter a new era of truly end-to-end (E2E) quality of service (QoS), service level agreements (SLAs) in the near future will be expected to include guarantees about the quality of security (QoSec) as well. The definition of ingredients of QoSec is currently being investigated: how to identify the security level required and to propose adaptive, dynamic and risk aware security solutions. 

Meanwhile, the evolution towards 6G systems is expected to introduce new means of reaching situational awareness by harvesting and interpreting the ``context'' of the communication, including, network tomography, nodes' constraints, the age of information, etc. 
Incorporating context awareness in QoSec is projected to allow handling more efficiently aspects related to identifying the risk or threat level and the required security level, particularly for applications with non-functional security requirements, such as autonomous vehicles, platooning, eHealth, etc. In this framework, incorporating security controls from the palette of physical layer security (PLS) can be particularly attractive due to their low computational complexity (relevant implementations are based on standard encoders) and their inherent ability to adapt to the transmission medium properties. The incorporation of PLS in 6G security requires indeed enhanced context awareness and can be particularly attractive for massive machine type communications (mMTC) and ultra-low latency use cases.  

In the rest of this article, we will begin in Section~\ref{secII} with a review of open security issues in 5G and research challenges ahead of 6G  and move on to presenting a roadmap 
to address these challenges 
in Section~\ref{secIII}. To illustrate some of the proposed ideas we outline viable solutions to address specific security vulnerabilities in 5G and 6G, along with a discussion of possible further directions in Sections IV and V, while conclusions are drawn in Section VI.

\section{Open 5G Security Issues and Security Research Challenges Ahead of 6G}
\label{secII}
Despite the strengthening of 5G security protocols with respect to previous generations, there are still open issues that have not yet been fully addressed, e.g., attacks under the generic umbrella of ``false base stations''. In parallel, in the path towards the 6G evolution, new security challenges arise as a result of drastic changes in key operation parameters: i) the E2E latency tolerance; ii) the sheer scale of networks in mMTC use cases and very large scale Internet of things (IoT); iii) the long lifespan of deployed IoT devices (notably sensors) that will need to be secured; iv) the wide variety of heterogeneous RF technologies involved; v) the accelerated steps taken towards bringing quantum computers to life. In the following, we provide a short review of open security issues in 5G and of some of the security challenges in the evolution towards 6G. This discussion provides the motivation for our proposal of context-aware security solutions for future generations of wireless, which will also be able to leverage the physical layer to provide flexible and adaptive security guarantees. 

\subsection{False Base Station Attacks}
The expression ``false base stations'' (FBS) describes impersonation attacks of genuine base stations. 
The topic is currently studied by the SA3 working group, documented in TR 33.809~\cite{1}. 
Typically in 5G an FBS is a ``man-in-the-middle'' (MitM) 
or a very stealthy jammer. A major vulnerability highlighted by FBSs is that the phases of entry into the network, which precede the enactment of the 5G security protocols, are particularly critical for many of the attacks described in TR 33.809. For example, attacks consisting in replaying modified versions of the broadcast channels can have disastrous consequences on all the terminals of a cell, hindering their connection to the network or forcing them to operate in a degraded mode. 
As a result, it is necessary to propose methods that allow the user equipment (UE) to determine whether a BS is legitimate, \textit{prior} to exchanging unauthenticated  messages. To this end, PLS could be used by incorporating the BS localization by a UE as a soft authentication factor.

\subsection{Security Challenges in Ultra Reliable and Low Latency Communications (URLLC)}

Critical ultra reliable low latency communications (URLLC) are typically used for industrial IoT (IIoT), vehicle-to-everything, and other applications requiring low latency and very high reliability. To achieve high reliability, a possible avenue is by increasing diversity, e.g., multiple parallel transmissions can be exploited. However, this consequently increases the ``attack surface'', while it might also impose more stringent constraints in terms of the speed of integrity checks. 
Overly aggressive latency targets could entail a new security architecture altogether. State-of-the-art proposals for fast authentication with use of implicit certificates or certificateless solutions can speed up authentication. Many open challenges for sub-millisecond delay constrained URLLC systems remain, with respect not only to authentication, but as well for the integrity and the confidentiality of both the control and data planes, as documented in~\cite{2}. 

\subsection{Jamming Attacks in mMIMO --- RF Resilience}
Although multiple input multiple output (MIMO) systems, including massive MIMO (mMIMO), make eavesdropping more difficult thanks to energy focusing, they nevertheless also introduce vulnerability points. Indeed, beamforming in mMIMO systems relies on accurate channel estimation. Pilots are transmitted in order to obtain the channel state information (CSI), which in turns allows precoding. If the CSI is not correctly estimated (e.g., because of interference or due to voluntary contamination by a jammer) the precoder will disperse the power, resulting in potential leakage and poor link quality. The later leads to service unavailability, giving rise to a denial of service (DoS) type of attack, as described in~\cite{3}. Similar attacks can also be launched at the medium access control (MAC) by tampering with the CSI reports sent by the devices. As a result, the beam management phase during network entry is vulnerable to RF jamming attacks. It is therefore crucial to have the means to detect, locate and neutralize jammers, or implement mitigation solutions.  

\subsection{Privacy}
Although 5G incorporates a set of measures to enhance privacy in terms of user identity (subscription) privacy, recent research
on user location privacy and user untraceability has shown that there are still many open issues, while 
the privacy guarantees are rather weak from an end-user perspective. 
The amount of personal data handled by future mobile networks will substantially increase 
Governmental agencies as well as adversarial entities have potentially a high interest in such data; future wireless networks have to be designed to ensure privacy without having to place trust in operators. 

\subsection{Post-Quantum Resilience }
A further challenge comes from quantum computing, which has seen significant progress after massive earlier investments. 
Since some of the most important cryptographic algorithms used in 5G are not quantum-resistant, the related protocols have to be redesigned involving post-quantum crypto algorithms. The national institute of standardization (NIST) is currently evaluating novel post-quantum crypto algorithms to replace currently used public key encryption schemes. Nevertheless, it is a common concern that quantum resistance will lead, at least in the immediate future, to an increase in terms of the complexity of the new cryptographic systems. For example bigger key sizes might pose a significant problem in practise. This could be especially challenging for URLLC and low-power / low-cost devices, further highlighting conflicting trends in future systems and the interplay between computational based crypto and real-time communication between low-end devices. 

\begin{figure*}
    \centering\begin{mdframed}[style=MyFrame]
\centering

    \includegraphics[width=0.8\textwidth]{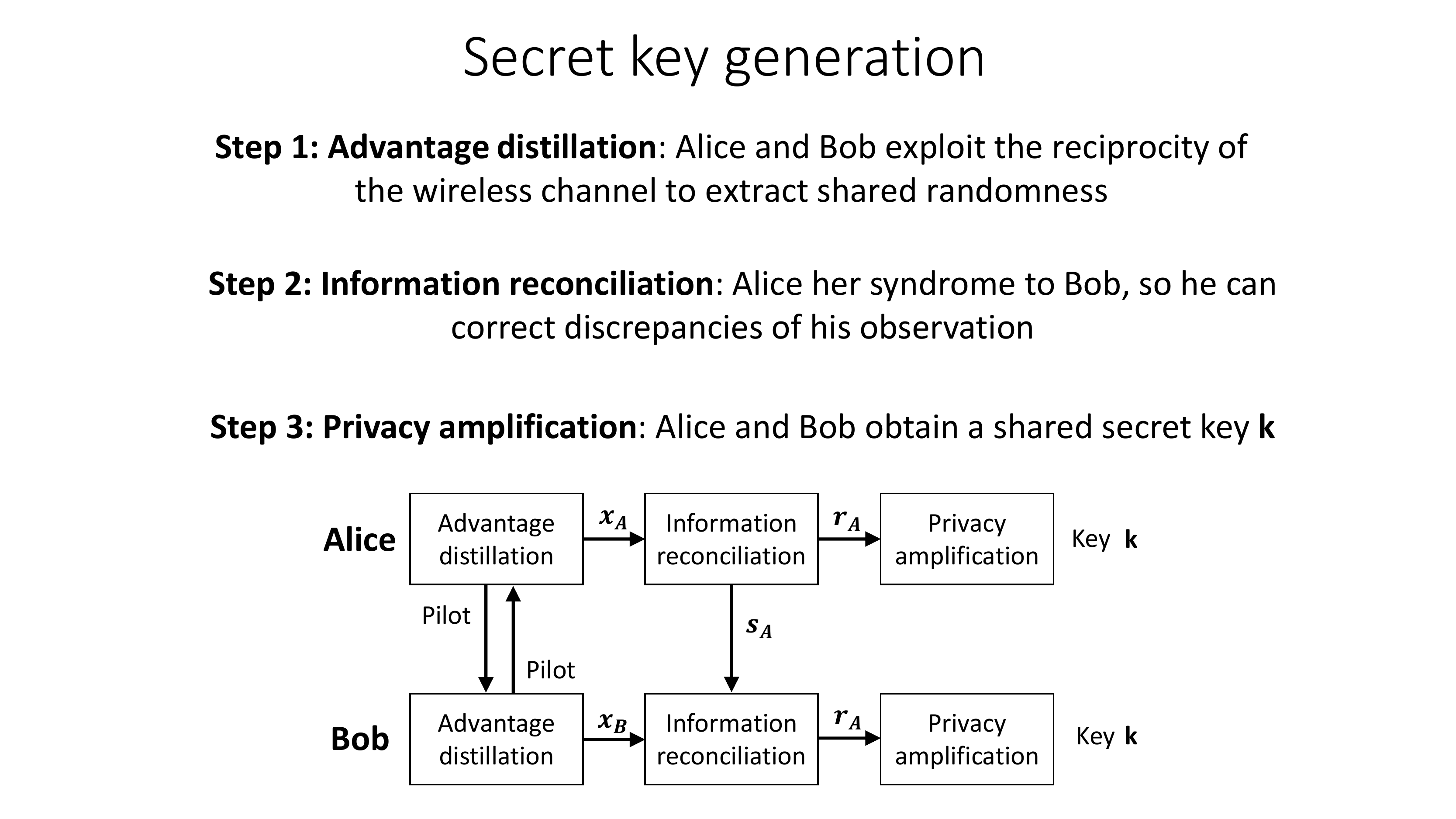}
    \includegraphics[width=0.6\textwidth]{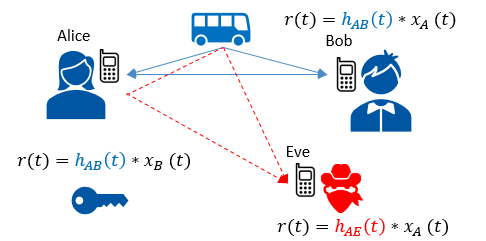}
    \caption{Distilling symmetric keys from wireless coefficients $h_{AB}$ in multipath channels, exploiting channel reciprocity during the channel's coherence time. The procedure comprises three phases, referred to as advantage distillation, information reconciliation and privacy amplification.}
    \label{figSKG}
    \end{mdframed}
    
\end{figure*}

\subsection{IoT security}
There are numerous security issues arising with the introduction of very large scale, long life, constrained IoT networks. 
Low-end, SIMless IoT devices, are unlikely to be able to support advanced security mechanisms, due to computing power, memory and -- probably most challenging -- energy consumption constraints. Although lightweight cryptography could help to address some of the challenges, such algorithms are currently not part of 5G and the development of lightweight post-quantum solutions is a recent field of research. 

Furthermore, the envisioned huge number (trillions) of very diverse IoT devices connected to the B5G network brings about big challenges in terms of information security management, but also is itself a security risk, as shown by the 2016 Mirai attack with a sever overall impact. 
In this aspect, decentralised intrusion / anomaly detection becomes important~\cite{5}. 

Another factor at play is that many IoT devices will typically have a very long lifespan ($>$10 years as opposed to 3 years for a laptop) and can be distributed in large geographical areas. 
It is difficult to guarantee that mass-produced, computationally and power constrained IoT devices will have a hardware capable of being updated with the necessary patches to resist all the threats that will arise in their lifetimes (e.g., post-quantum resistance).


\section{6G as an  Enabler to Context Aware QoSec Leveraging PLS}
\label{secIII}

Even though 6G is still some years away from standardization, consensus is growing on its likely evolution path, 
briefly outlined in the following. 

 \begin{figure}[t]
    \centering
    \includegraphics[width=0.5\textwidth]{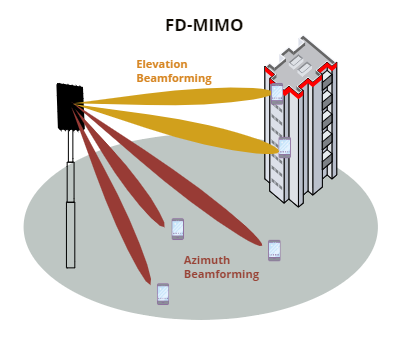}
    \caption{mMIMO full dimension beamforming
    } \label{fig:mMIMO}
\end{figure}

\begin{itemize}
   \item  \textit{\textbf{Higher frequencies and bandwidth}}: Continuing the evolution seen in the previous generations, 6G will make use of ever higher carrier frequencies and bandwidth, moving towards frequencies above 100 GHz, which allows the allocation of bandwidths larger than 1 GHz. The large bandwidth may increase the observable channel entropy in the frequency domain, which can potentially be exploited in PLS secret key generation (SKG) from wireless coefficients~\cite{7}, whose principal mechanisms are depicted in Fig.~\ref{figSKG}. 
   
 Additionally, in mmWave and SubTHz systems beamforming with pencil-sharp beams becomes both a possibility, because of the smaller area occupied by antenna arrays, and, a necessity, because of the need to compensate for the higher channel attenuation. Highly directive beamforming can then reduce eavesdropping opportunities, as depicted in Fig. \ref{fig:mMIMO}, while similar opportunities exist for visible light communications (VLC). Thus, in 6G, massive MIMO (mMIMO) could offer a viable application scenario for the wiretap channel. 
 

 \item   \textit{\textbf{Integrated sensing and communications}}: In addition to high-resolution image, video and sound, among other possible sensing data, which can be transmitted through mobile communication networks, radar sensing is likely to be an integral part of future wireless systems~\cite{8}, reusing the same spectrum and waveform as communications. These new capabilities along with centimetre-level localization precision will allow the network to have a better understanding of the surroundings and gain situational awareness, i.e., understanding of the context of communication. On the other hand, this raises other security issues, as the sensing data themselves may be subject to tampering by attackers. Their integrity must be assured. As a result, trustworthiness of sensing and communications is expected to be a key performance indicator of 6G systems.

 \item \textit{\textbf{Learning at the wireless edge and native AI}}: Centralized machine learning, which processes data centrally using cloud-based computing, is subject to critical security challenges, e.g., a single point of failure and the vulnerability of data during backhaul. Moreover, it might not be suitable for real-time applications, due to the capacity and latency requirements  resulting from centralized data aggregation and processing. Thus, decentralized ML solutions are becoming increasingly important, e.g., federated learning, in which data are in principle processed locally at end-user devices where they are collected. While such distributed ML solutions can serve as enabling technologies for 6G mobile edge networks, they also introduce vulnerabilities such as the leakage of private information through learned model parameters, exposure to malicious end-user devices and adversarial training examples.
\end{itemize}
   
These anticipated 6G features provide novel opportunities to address the security and privacy challenges outlined in Section~\ref{secII}, allowing for the security architecture of 6G networks to be built around automation. Following the principles of multilateral security, 
the system should understand the security goals of the entities involved and should adapt the security controls accordingly based on contextual information, harvested from the novel 6G features. To this end, we need a set of building blocks: 

\begin{enumerate}
\item[i)]  Quantify security in the QoSec framework, i.e.,  the ability to express the desired and actual ``level of security'';
\item[ii)] Context awareness at the wireless edge with the aid of sensing and AI;
\item[ii)] New, adaptive security controls, incorporating PLS;
\item[iv)] Automation in the form of a ML/AI based security orchestrator. 
\end{enumerate}
In the following subsections we discuss in further detail some of these necessary building blocks.

\subsection{Quantifying Security: Quality of Security (QoSec)}
Similar to QoS definitions (e.g., [ITU-T E.800]), QoSec is the totality of characteristics of a service that bear on its ability to satisfy stated and implied security needs of the user. QoSec is able to provide different security guarantees, in response to the security needs of different use cases and related slices of the network, reflecting on the DiffServ QoS paradigm. A central aspect related to QoSec is to identify how to make the security level and its implementation adaptive: how to automatically identify the right QoSec and the right combination of crypto schemes (encryption, integrity, authentication primitives), as well as how to incorporate these flexibly in security protocols. 

Thereby, adaptivity can happen at different levels: for a fixed cryptographic strength (e.g., 256-bit symmetric block ciphers considering quantum-resistant) and a fixed attacker model (e.g., ``zero trust'', i.e., minimal (trust) assumptions regarding all involved entities) we can adapt the specific cryptographic algorithms and protocols that are used~\cite{9}. On the other hand, we could also adapt the desired cryptographic strength or the considered attacker model based on contextual information. In future security protocols varying levels of trustworthiness (e.g., as defined by NIST in SP800-53 Rev. 4) are envisioned through the use of security control baselines. Note that these are developed based on a number of general assumptions, including common environmental, operational and functional considerations, giving rise to the question of context awareness in security. In Section IV we discuss in detail how PLS can be leveraged to develop adaptive security controls.


\subsection{Context Awareness at the Wireless Edge: The Role of Sensing and AI}

The opening of the THz spectrum will provide new ``sensing'' capabilities to 6G devices, such as high-definition imaging and frequency spectroscopy. 
Unique opportunities arise for reaching context awareness through the processing of sensing information with both centralized and edge AI; in turn, context awareness is key for trust building and for predicting reliability, i.e., QoSec can be driven by context awareness. Incorporating context awareness in security controls amounts to being able to provide answers -- with the aid of AI~-- to the following open-ended questions: 
\begin{enumerate}
 \item[1]  \textit{\textbf{How to extrapolate the threat level from context}}: PHY layer inputs, particularly in the form of sensing information including the location of a node, the time of communication, the ambient temperature, etc., carry important contextual information, directly related to semantics. We can envision AI multi-modal fusion of sensing information to obtain an enhanced evaluation of the threat level. In very demanding scenarios such as platooning, this approach might help provide a viable route to develop anomaly detection solutions for highly dynamic, seemingly chaotic, networks.

 \item[2]   \textit{\textbf{How to use context to identify the security level required}}: We need to take steps towards defining new metrics describing the criticality of the particular data exchanged and furthermore, how valuable they are considered from an adversarial point of view. This can be thought of as the analogous of defining the priority level in QoS. 

 \item[3]  \textit{\textbf{How to match security levels to security schemes}}: After defining the security level with rapport to the context of communication, the next question is how to map this to an actual set of algorithms and security schemes. Two approaches emerge that can possibly be used jointly: i) crypto based approaches, in which the strength of crypto systems is, roughly speaking, related to the lengths of the keys (after the right transformations are accounted for); ii) PLS approaches, in which the wireless channel and the hardware are used as sources of uniqueness (for authentication) and / or entropy for confidentiality purposes (e.g., for SKG) [7]. Next, we delve into the potential use of PLS in 6G and discuss how PLS is inherently adaptive and can be enabled by context awareness.
\end{enumerate}

\section{QoSec Adaptive Security Controls: The Role of Physical Layer Security in 6G}
In the past years, PLS~\cite{10,11} has been studied and indicated as a possible way to emancipate networks from classic, complexity based, security approaches~\cite{12}. PLS is based on the premise that we can complement some of the core security functions, exploiting both the communication radio channel and the hardware as sources of uniqueness or of entropy. 

It is usually this latter aspect of PLS that is considered in the literature, around the concept of the secrecy capacity and of the SKG capacity~\cite{13}. In this framework, 
PLS leverages the physical properties of the radio channel, namely diffusion, superposition and reciprocity, to create opportunities for secure data transmission in the presence of eavesdroppers in the channel.  These properties can be exploited in a variety of ways, including taking advantage of independent fading between legitimate users and eavesdroppers, the use of multiple-antennas or relays and the injection of artificial noise to create secure degrees of freedom. 

 \begin{table}[t]
\caption{Minimum number of antennas $N_t$ required at the BS of a downlink MISO network in the presence of a single antenna eavesdropper for a target secrecy outage probability $P_{so}$, using eq. (11) of \cite{MISO}.}
\centering
\begin{math}
\begin{array}{|c|c|c|c|c|}
\hline
   &  P_{so}=10^{-1} & P_{so}=10^{-3} & P_{so}10^{-5} & P_{so}10^{-10}\\
    \hline
    \hline
   \begin{array}{l} 
   \alpha_e / \alpha_s=1/4 \\
    \alpha_e / \alpha_s=1/2 \\
    \alpha_e / \alpha_s=1 \\ 
    \end{array}
    & 
     \begin{array}{l}
    1 \\
    2 \\
    9 \\
    \end{array} 
    & 
     \begin{array}{l}
    3 \\
    10 \\
    952 \\
    \end{array}
        & 
    \begin{array}{l}
    7 \\
    37 \\
    - \\
    \end{array}
    &
     \begin{array}{l}
    21 \\
    698 \\
    - \\
    \end{array}\\
        \hline
\end{array}
\end{math}
\label{ta:MU-MIMO}
\label{table2}
\end{table}

 \begin{table}[t]
 \caption{Maximum eavesdropper density $\lambda_e$ to achieve a target secrecy outage probability $P_{so}$ in an UAV network using eq. (17) in \cite{UAV}. $\lambda_u$ denotes the density of legitimate nodes, UAV at a height of $H=10~m$.  }
\centering
\begin{tabular}{|c|c|c|c|c|}
\hline
     &$P_{so}=10^{-1}$ & $P_{so}=10^{-3}$ &  $P_{so}=10^{-5}$  & $P_{so}=10^{-10}$\\
    \hline\hline
    $\lambda_u =10^{-3}$ & $ 1$ & $ 10^{-1}$ & $10^{-3}$ &$10^{-8}$\\
     \hline
     $\lambda_u =10^{-2}$ & $1$ & $10^{-2}$ & $10^{-4}$ & $10^{-9}$\\
    \hline
\end{tabular}
\label{ta: UAV}
\end{table}

In the celebrated wiretap channel model introduced by Wyner in 1975, the adversarial link is degraded with respect to the main link, i.e, legitimate users do not share a secret bur have a link quality advantage; whenever this can be substantiated, the existence of wiretap codes that can ensure asymptotically both reliability in the reception of a confidential message by a legitimate receiver and negligible information leakage to an eavesdropper has been demonstrated. Furthermore, by adjusting network / system parameters, different secrecy outage probabilities -- potentially corresponding to different QoSec levels -- can be attained. We  illustrate the underlying ideas in uses cases in which the wiretap channel is used to convey securely symmetric  secret keys in hybrid PLS-crypto systems. In this case, very low secrecy rates can be targeted as a single key of $256$ bits can be used to encrypt up to Gigabytes of data, e.g., when wiretap coding is used jointly with modern ciphers such as the advanced encryption algorithm (AES) in Galois counter mode (GCM), then negligible secrecy rates in the order of $10^{-7}$ could be sufficient.  Under this assumption, we illustrate system design parameters to achieve positive secrecy rates in two scenarios: i) first we evaluate the minimum number of antennas at a BS in Table \ref{ta:MU-MIMO} for MISO channels using the results in \cite{MISO}; ii) secondly, in Table \ref{ta: UAV} the maximum eavesdropper density is evaluated for unmanned aerial vehicle (UAV) networks based on \cite{UAV}.

In Table I,  notice the critical role of the relative quality of the adversarial versus the legitimate link, captured in ratio of the corresponding large scale fading coefficients, denoted by $\alpha_e$ and $\alpha_s$ respectively. When the legitimate user is much closer to the BS than the adversary, a secrecy outage probability as low as $10^{-10}$ can be attained with the use of only $21$ antennas.   On the other hand, due to line-of-sight in UAV communications, the maximum eavesdropper density for the same secrecy outage probability is $\lambda_e=10^{-8}$ when the legitimate node density is $\lambda_u=10^{-3}$ and the UAV is 10 m above ground. Only by reducing the secrecy outage probability, i.e., by reducing the target QoSec level, can the maximum eavesdropper density be increased. These two examples demonstrate that context awareness is necessary for the correct employment of PLS; in the MISO setting, proximity to the access point is critical, in the UAV example node density plays a major role. These two examples further show that in a given context, PLS can be used to achieve potentially a subset of QoSec levels, articulated around secrecy outage probabilities.

Another important point for the use of wiretap codes in 6G arises with respect to low latency systems using short packets. In the finite blocklength wiretap channel it is not possible to achieve zero information leakage as in the asymptotic regime \cite{15}. Rather, a small quantity $\delta$ is introduced as a guaranteed upper bound in terms of information leakage. We envision that in low-end IoT networks, in which potentially low security QoSec levels could be acceptable, tolerating a maximum leakage rate of $\delta$ could be a possible route to provide privacy guarantees.

 \begin{figure}
        \centering
        \includegraphics[width=0.5\textwidth]{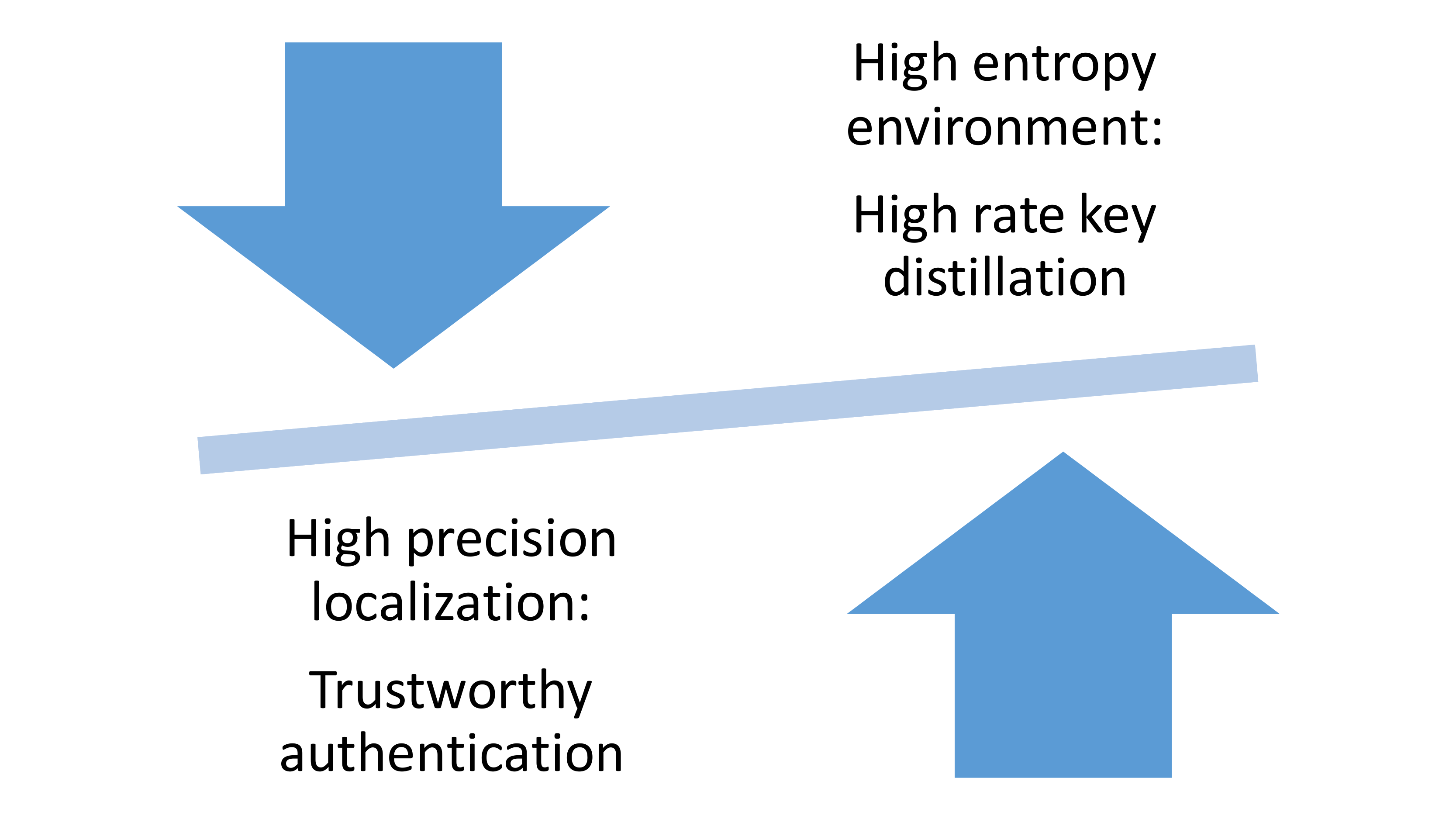}
        \caption{The wireless channel can act as a source of entropy or as a source of high precision localization and positioning for authentication. }
        \label{figUniqVsAuth}
        
    \end{figure}

Furthermore, as mentioned earlier, the use of PLS will profit from the pencil-sharp beams likely to be available in 6G~\cite{14}, as they will make eavesdropping very difficult by attackers not located in the beam direction, while the same is true for visible light communications~\cite{12}. 
Additionally, the high bandwidth may provide enough entropy to help the generation of high-rate secret keys~\cite{7}. SKG schemes from wireless coefficients are probably the most mature of all PLS technologies. However, context awareness is critical for the incorporation of SKG in 6G systems. In particular, as the line-of-sight conditions and the channel quality change, there is a clear trade-off between the use of the wireless fading for high-precision localization, which is key  for the PLS authentication discussed next, and as the means to distill entropy for key agreement, confidentiality and integrity schemes, showcased in Fig.~\ref{figUniqVsAuth}. This unique setting can only be exploited with enhanced monitoring of the wireless channel and of the context in general, confirming once more that context awareness is indeed an enabler for PLS.

With respect to user authentication, we can leverage the PHY by using RF fingerprinting and high-precision localization as soft authentication factors.
It is worth mentioning that many new features of future networks, like low-latency control loops, sensor fusion or simultaneous localization and mapping (SLAM) will require only local communications, not involving the core network. These can be made more secure and agile if PLS is employed, alleviating the need for network-based centralized security. In this context, PLS enabled by ML can be used for intelligent PHY authentication in dynamic and complex 6G environments such as IoT networks. Thanks to the ability of ML techniques to learn and capture statistics of complex features, we can achieve low-cost, continuous, highly reliable, model-independent, and context-aware authentication, e.g., leveraging localization and RF fingerprinting. 
To enhance the reliability of such authentication mechanisms, the trustworthiness of the observed and estimated attributes needs to be monitored, accounting for context.

Finally, in terms of device authentication, it is further possible to leverage ``hardware fingerprints'' in the form of physical unclonable functions (PUFs), as an authentication factor in multi-factor authentication protocols. PUFs rely on the use of Wyner-Ziv reconciliation approaches to offer measurable re-usability of the hardware fingerprint. Combining various PLS technologies, hybrid PLS-crypto systems can be built around the ideas of zero-round trip time (0-RTT) protocols and / or authenticated encryption [13], offering further tools to develop fast authentication schemes at PHY, potentially exploiting multiple authentication factors.

   \begin{table*}[t]
\caption{Roadmap of Solutions for 5G / 6G Security Challenges} \label{tb:1}
\centering\renewcommand{\arraystretch}{1.6}
  \centering
\begin{tabular}{|l|l|}

\hline
{Security Challenge / Scenario}  & {Recommended techniques (with $^\ast$ we denote PLS / PHY solutions)} \\ 
\hhline{|=|=|}
  False Base Station
Attacks &  
\begin{tabular}{l}
$^\ast$ Intelligent PHY authentication using RF fingerprinting and localization of BS from UE (inverse localization) \\
$^\ast$ Pre-shared keys established / distributed with SKG
\end{tabular} \\
\hline
Low Latency Communications & 
\begin{tabular}{l}
$^\ast$ Fast authentication using PUFs and RF fingerprinting as early authentication factors \\
$^\ast$ Short packet secrecy encoding\\
$^\ast$ Short blocklength Slepian Wolf and Wyner Ziv reconciliation decoders (for SKG and PUFs)
\end{tabular}\\
\hline
Jamming Attacks in mMIMO --- RF Resilience & 
\begin{tabular}{l}
$^\ast$ Spectrum sensing, channel charting, channel learning \\
$^\ast$ Advanced modulation and coding \\
$^\ast$ Intrusion detection at PHY\\
$^\ast$ Covert communications / low probability of detection 
\end{tabular}\\
\hline
Privacy & 
\begin{tabular}{l}
   - Context aware choice of pseudonymity, partial identities\\
 - Contextual aware integrity to detect and mitigate violations\\
   - Context aware appropriateness and distribution   \end{tabular}\\
\hline
Post-Quantum Resilience & 
\begin{tabular}{l} 
$^\ast$ PLS is information theoretic secure \\
$^\ast$ Long symmetric encryption keys using channel-based key generation\\
$^\ast$ Hybrid crypto-PLS schemes
 \end{tabular}\\
 \hline
 
Low-cost IoT devices &
\begin{tabular}{l}
$^\ast$PLS is lightweight, secrecy encoders, SKG, PUfs, etc. \\
Awareness of low-cost / low-security IoT devices for appropriate isolation in a dedicated network slice
\end{tabular}\\
\hline
Huge Number of IoT devices &
\begin{tabular}{l}
- Contextual understanding to automatically select appropriated QoSec\\
- Adaptive and automatic security controls removing the burden to manually configure and monitor all the IoT devices\\
$^\ast$ PLS as a scalable technique  for key management and distribution
\\
$^\ast$ PLS as adaptive security scheme
\end{tabular}\\
\hline
Long-term IoT security & 
\begin{tabular}{l}
- Awareness of a decrease over time in QoSec and trusthwortiness \\
- Automatic adoption of the overall security controls and policies\\
- Context aware access control, e.g., excluding untrustworthy devices from the network or reduction of (access) rights
\end{tabular}\\
\hline
\end{tabular}
\label{ta: roadmap}
\end{table*}

\section{Discussion and Proposed Roadmap}

Looking at the broader picture, down the path towards 6G, novel security challenges and  opportunities arise. Among the challenges, noteworthy are issues related to vulnerabilities in the initial entry phases of a node in a network (before the enactment of the 5G security protocols), the massive number of low-end and heterogeneous IoT devices, sub-millisecond delay constraints for critical IoT use cases, etc., while offering post-quantum security guarantees and addressing issues of privacy. On the other hand, 6G is expected to be the first generation of wireless to offer edge- and device-level intelligence, leveraging novel sensing capabilities and the extensive use of ML. 
The incorporation of context awareness in 6G security protocols can propel the introduction of disruptive new technologies to provide flexible and adaptive security guarantees, based on an on-line evaluation of the security threat level. 

It is in this context that PLS technologies can be truly exploited; PLS can be realised only with provably trustworthy monitoring and understanding of the communication environment and communication medium in 6G. In applications such as the IoT, PLS emerges as a very competitive candidate to be used in context-aware, flexible and adaptive security controls, both for authentication as well as for confidentiality schemes. While PLS might not, at least in the near future, be incorporated in zero-trust security protocols, it does provide a viable alternative to securing massive and ultra-low latency networks with relaxed security guarantees, as a competitive candidate for emerging QoSec approaches that will cut across all layers of the network stack.

PLS offers notable advantages. Firstly, it is inherently adaptive; by adjusting the target secrecy rate or secret key rate, one can adapt related secrecy outage probabilities, offering a flexible framework with respect to adaptive security controls. Furthermore, 
PLS can provide information-theoretic security guarantees using lightweight mechanisms (e.g., using Polar or low density parity check (LDPC) encoders)  as opposed to computationally expensive cryptographic schemes. Thus, such approaches are suitable for low complexity IoT devices and for networks with light or no infrastructure, either as stand-alone best-effort security mechanisms or as complements to more traditional methods.

To exemplify some of the points made previously, in Table \ref{ta: roadmap} we present a roadmap on how to address the security challenges listed in Section~\ref{secII} and how PLS fits into this picture. We want to emphasize that the presented ideas are still just parts of the puzzle and have to be embedded in a much more holistic approach, which, besides additional technical means, has to incorporate organisational, regulatory, economical -- and not to forget, standardisation -- aspects.

\section{Conclusions}
Unarguably, 5G security enhancements present a big improvement with respect to LTE. However, as the complexity of the application scenarios increases with the introduction of novel use cases, notably URLLC and mMTC, novel security challenges arise that might be difficult to address using the standard paradigm of complexity based classical cryptographic solutions.  At the same time, in the longer 10-year horizon novel security concepts based on “trust models” and risk-based, adaptive identity management and access control will come to life, enabled to a large extend by AI.  To allow for flexible QoSec, the development and integration of security controls at all layers of the communications system is envisioned. 

In this framework,  PLS  is  being  considered  as  a  possible  way  to  emancipate networks from classical, complexity based, security approaches.  
With respect to authentication, PUFs, wireless fingerprinting / localization, combined with more classical approaches, could also enhance authentication and key agreement (AKA) in demanding scenarios. In parallel, THz communications will rely upon setting up highly directional beams, potentially providing a concrete scenario for the wiretap channel. Furthermore, with the opening up of higher frequency bands in 6G, the opportunity to harness entropy in the frequency domain can be exploited in SKG protocols. As a general direction, context awareness, enabled by enhanced sensing and AI capabilities anticipated in 6G,  can allow introducing  disruptive tools for providing adaptive QoSec based security guarantees, tailored to the context of the communication, incorporating PLS security controls.


\vspace{1cm}

\noindent\textbf{Arsenia (Ersi) Chorti} is a Professor at the École Nationale Supérieure de l'Électronique et de ses Applications (ENSEA), ETIS Lab UMR 8051 and a Visiting Scholar at Princeton and Essex Universities. Current research topics include physical layer security for 5G / 6G and IoT, anomaly detection, machine learning for communications, NOMA and scheduling. She is a Senior IEEE Member, member of the IEEE INGR on Security, the Competitive Pole Systematic and of the PhD Thesis GdR ISIS Award Committee in France. Since October 2021 she is chairing the IEEE Focus Group on Physical Layer Security.

\vspace{1 cm}
\noindent\textbf{André Noll Barreto}
André Noll Barreto (Senior Member, IEEE) received an M.Sc. from PUC, Rio de Janeiro, Brazil, in 1996, and a Ph.D. from the TU Dresden, Germany, in 2001, both in Electrical Engineering. He held several positions with academia and industry, at IBM Research – Zurich, Claro, Nokia Technology Institute/INDT, Universidade de Brasília  and at his own start-up, Ektrum. He joined the Barkhausen Institut, in Dresden, Germany, in 2018. His current research interests are in Physical Layer Security and Joint Communications and Sensing, and in the development of open-source simulators for 6G systems.

\vspace{1cm}
\noindent\textbf{Stefan Köpsell}
Stefan Köpsell is the leader of the Secure and Privacy-Respecting Data Processing Group at the Barkhausen Institut (since 2019) and an acting professor for the Chair of Privacy and Data Security at TU Dresden. He has a PhD in computer science (2010) from TU Dresden, which he joined 2000. His research focus includes security in and by distributed systems, privacy and anonymous communication as well as IoT security. He has published more than 75 articles in refereed journals and conferences.

\vspace{1cm}
\noindent\textbf{Marco Zoli}
Marco Zoli received the Ph.D. degree in radio channel characterization for future wireless networks and applications from the University of Bologna, Italy, in 2018. He joined the Barkhausen Institut, Dresden, Germany, in 2019, as Research Associate, working on physical layer security. His main research interests included antennas, telecommunications, wireless technologies, security and privacy, numerical simulations, and open science. Since October 2021 he has joined Xilinx Dresden GmbH, an AMD company.

\vspace{1cm}
\noindent\textbf{Marwa Chafii}
Marwa Chafii received her Ph.D. degree in electrical engineering in 2016 from CentraleSupélec, France. She joined TU Dresden, Germany, in 2018 as a group leader, and ENSEA, France, in 2019 as an associate professor. Since 2021, she is an associate professor at New York University (NYU) Abu Dhabi, and NYU WIRELESS.  She received the prize of the best Ph.D. in France in the fields of Signal, Image and Vision, and she has been nominated in the top 10 Rising Stars in Computer Networking and Communications in 2020.  

\vspace{1cm}
\noindent\textbf{Philippe Sehier}
Philippe Sehier is a department head at Nokia Bell Labs France. His current research interests are in the area of 5G and 6G access network, and more specifically, the radio interface. After a career in research and development in civil and military satellite and terrestrial transmissions, he joined the mobile business of Alcatel Lucent in 2001 where he held various product management and marketing positions for 3G, 4G and currently 5G and 6G. He graduated from the Ecole Supérieure d'Electricité in 1984. He is the author of numerous publications and patents in the fields related to radio transmission.

\vspace{1cm}
\noindent\textbf{Gerhard Fettweis}
Gerhard P. Fettweis, F’09, earned a Ph.D. under H. Meyr at RWTH Aachen in 1990. After a postdoc at IBM Research, San Jose, he joined TCSI, Berkeley, USA. Since 1994 he is Vodafone Chair Professor at TU Dresden, Germany. Since 2018 he additionally heads the new Barkhausen Institute. 2019 he was elected into the DFG Senate (German Research Foundation). He researches wireless transmission and chip design, coordinates 5G++Lab Germany, spun-out 17 tech and 3 non-tech startups, and is member of the German Academy of Sciences (Leopoldina), and German Academy of Engineering (acatech). 

\vspace{1cm}
\noindent\textbf{H. Vincent Poor} is the Michael Henry Strater University Professor at Princeton University, where his interests include information theory, machine learning and network science, and their applications in wireless networks, energy systems and related fields. Among his publications in these areas is the forthcoming book Machine Learning and Wireless Communications (Cambridge University Press). A member of the U.S. National Academies of Engineering and Sciences, he received the IEEE Alexander Graham Bell Medal in 2017.

\end{document}